\makeatletter
\@ifundefined{@parse@version@dash}{%
\def\@parse@version#1{\@parse@version@0#1}
\def\@parse@version@#1/#2/#3#4#5\@nil{%
\@parse@version@dash#1-#2-#3#4\@nil}
\def\@parse@version@dash#1-#2-#3#4#5\@nil{%
  \if\relax#2\relax\else#1\fi#2#3#4 }
}{}
\makeatother

\documentclass[%
 reprint, onecolumn,
superscriptaddress,
 amsmath,amssymb,
 aps,
]{revtex4-2}

\usepackage{graphicx}
\usepackage{wrapfig}
\usepackage{hyperref}
\usepackage{dcolumn}
\usepackage{bm}

\usepackage{soul}
\emergencystretch=\maxdimen
\begin{document}


\title{Evidence for a QCD accelerator in relativistic heavy-ion collisions}

\author{L.~C.~Bland}\affiliation{Temple University, Philadelphia,
  Pennsylvania  19122, USA}
\author{E.~J.~Brash}\affiliation{Christopher Newport University, Newport News, Virginia 23606, USA}
\author{H.~J.~Crawford}\affiliation{University of California,
  Berkeley, California  94720, USA}
\author{A.~Drees}\affiliation{Brookhaven National Laboratory, Upton,
  New York 11973, USA}
\author{J.~Engelage}\affiliation{University of California,
  Berkeley, California  94720, USA}
\author{C.~Folz}\affiliation{Brookhaven National Laboratory, Upton,
  New York 11973, USA}
\author{E.~Judd}\affiliation{University of California,
  Berkeley, California  94720, USA}
\author{X.~Li}\affiliation{Los Alamos National Laboratory, Los Alamos, New Mexico 87545, USA}
\author{N.~G.~Minaev}\affiliation{National Research Center ``Kurchatov Institute'', Institute of
  High Energy Physics, Protvino 142281, Russia}
\author{R.~N.~Munroe}\affiliation{Christopher Newport University, Newport News, Virginia 23606, USA}
\author{L.~Nogach}\affiliation{National Research Center ``Kurchatov Institute'', Institute of
  High Energy Physics, Protvino 142281, Russia}
\author{A.~Ogawa}\affiliation{Brookhaven National Laboratory, Upton,
  New York 11973, USA}
\author{C.~Perkins}\affiliation{University of California,
  Berkeley, California  94720, USA}
\author{M.~Planinic}\affiliation{University of Zagreb, Zagreb, HR-10002, Croatia}
\author{A.~Quintero}\affiliation{Temple University, Philadelphia,
  Pennsylvania  19122, USA}
\author{G.~Schnell}\affiliation{University of the Basque Country UPV/EHU, 48080 Bilbao \&
  IKERBASQUE, Basque Foundation for Science, 48009 Bilbao, Spain}
\author{G.~Simatovic}\affiliation{University of Zagreb, Zagreb, HR-10002, Croatia}
\author{P.~Shanmuganathan}\affiliation{Kent State University, Kent, Ohio  44242, USA}\affiliation{Brookhaven National Laboratory, Upton,
  New York 11973, USA}
\author{B.~Surrow}\affiliation{Temple University, Philadelphia,
  Pennsylvania  19122, USA}
\author{A.~N.~Vasiliev}\affiliation{National Research Center ``Kurchatov Institute'', Institute of
  High Energy Physics, Protvino 142281, Russia}\affiliation{National Research Nuclear University
  ``Moscow Engineering Physics Institute'', Moscow, Russia}


\vspace{0.1cm}

\date{\today}

\begin{abstract}

We report measurements of forward jets produced in Cu+Au collisions at
$\sqrt{s_{NN}}=200$ GeV at the Relativistic Heavy Ion Collider.  The
jet-energy distributions extend to energies much larger than expected
by Feynman scaling.  This constitutes the first clear evidence for
Feynman-scaling violations in heavy-ion collisions.  Such high-energy
particle production has been in models via QCD string interactions,
but so far is untested by experiment.  One such model calls this a
hadronic accelerator.  Studies with a particular heavy-ion event generator
(HIJING) show that photons and mesons exhibit such very high-energy
production in a heavy-ion collision, so {\it QCD accelerator}
appropriately captures the physics associated with such QCD string
interactions.  All models other than HIJING used for hadronic
interactions in the study of extensive air showers from cosmic rays
either do not include these QCD string interactions, or have smaller
effects from the QCD accelerator.

\end{abstract}
  
\maketitle

\section{Introduction}
\begin{figure}
  \centering
  \includegraphics[trim=0.5cm 0.1cm 0.5cm 0.5cm, clip, width=0.6\linewidth]{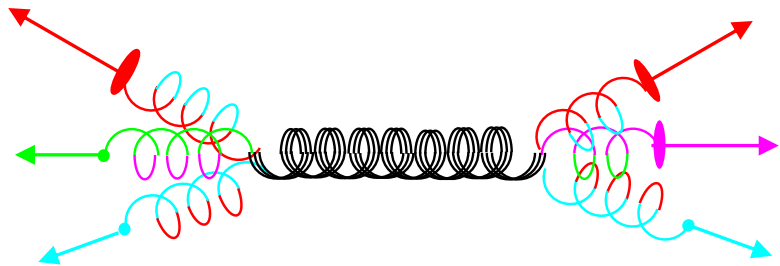}
  \caption{A schematic of a QCD accelerator.  Color charge is
    separated from anti-color charge in a hadronic collision.  A QCD
    string, represented in the schematic as a gluon, is stretched
    between the separated charges.  The schematic shows that three
    nearby QCD strings {\it fuse}, as they can because the QCD fields
    carry color charge.  The fused string can produce particles with
    longitudinal momentum component $>\sqrt{s_{NN}}/2$.}
  \label{QCD-accel}
\end{figure}

Quantum chromodynamics (QCD) describes the interaction of quarks and
gluons (collectively known as partons) due to their color
charges. Color is confined in hadrons such as protons and neutrons
(collectively known as nucleons), from which atomic nuclei are built.
Due to color confinement, partons are studied in high-energy
collisions characterized by total center-of-mass energy $\sqrt{s}$.
When hadrons collide at high
energy, most particles are produced by partonic interactions,
consisting of a parton from one hadron interacting with a parton from
the other hadron.  Scattered partons manifest themselves as jets,
which are sprays of particles.  Scattered partons are connected by QCD
field lines to other color charges.  The quanta of QCD fields are
gluons, which carry color charge.  A QCD string is a flux tube between
color charges arising because gluons interact with other gluons.
Under appropriate conditions, QCD strings can interact with other
strings (as shown schematically in Fig. \ref{QCD-accel}) resulting in the
theoretical prediction that forward particle production can extend to energies
higher than expected for pairwise parton interactions \cite{Ar961,Ar962}.

When the colliding hadrons are heavy ions (HI), the complexity of the
collision increases.  The conventional view of a HI collision is that
it involves multiple instances of a nucleon from one incoming ion
interacting with a nucleon from the other incoming ion.  These
nucleons collide with center-of-mass energy $\sqrt{s}$ taken as the
center-of-mass energy of the colliding heavy ions ($\sqrt{s_{NN}}$).
It is typical to scale the incoming
momentum by the number of nucleons in each HI and compute
$\sqrt{s_{NN}}$ from the scaled momenta.  
Partons from these nucleons then scatter.  The complexity of the HI
collision is from the superposition of all these scatterings and from
strong final-state interactions of the scattered partons.  Jets, or
their surrogates, that are produced in heavy-ion collisions near midrapidity lose energy when
traversing the hot and dense medium produced in the
collision \cite{STAR02,PHEN02,CMS12,ATLAS13,ALICE15}.  This medium is a
quark-gluon plasma \cite{BRS18}.

\begin{figure}
  \centering
  \includegraphics[width=0.5\linewidth]{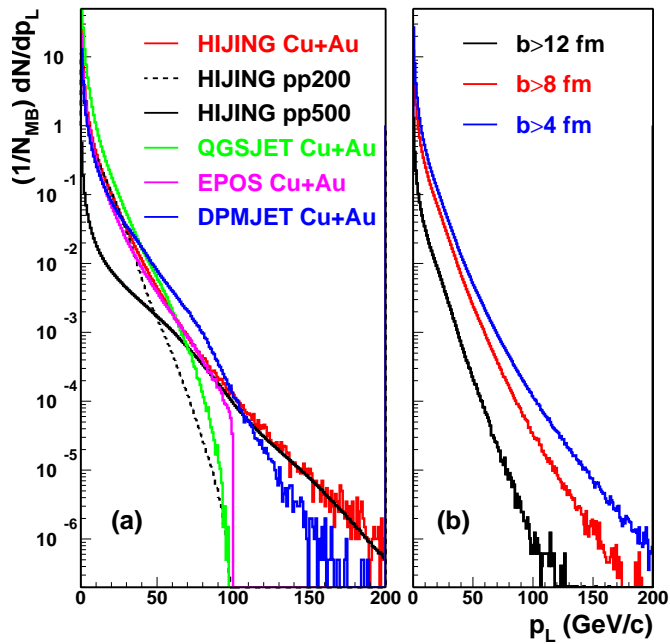}
  \caption{(a) Model predictions for the distribution of
    longitudinal momenta for positive pions produced in a hadronic
    collision normalized to the number of minimum-bias collisions. The Cu+Au predictions are at
    $\sqrt{s_{NN}}=200$ GeV and the p+p predictions are at the $\sqrt{s}$
    noted in the key.  (b) HIJING \cite{WG91} predictions for the
    dependence of the $p_L$ distributions on the impact parameter for
    Cu+Au collisions at $\sqrt{s_{NN}}=200$ GeV.}
  \label{models}
\end{figure}

Particles are produced in hadronic collisions as the QCD strings break with a variety of values
of transverse ($p_T$) and longitudinal ($p_L$) momentum.  In general,
most hadronic scattering experiments focus on the $p_T$ dependence.  For a
given $\sqrt{s}$, hadrons produced in the forward direction are found
to have a limiting value of $p_L$ of $\sqrt{s}/2$, as seen for model
calculations of p+p collisions at $\sqrt{s}=200$ GeV in
Fig.~\ref{models}a.  Feynman scaling \cite{Fe69} is the expectation the $p_L$
distribution will have a similar shape in the approach to the limiting
value for different hadronic collisions.
The scaling variable is $x_F=2p_L/\sqrt{s}$.  It is expected that
$x_F<1$.  In general, studies of only the $p_T$ dependence can not
determine the appropriate nucleon-nucleon (NN) $\sqrt{s}$, and assume
$\sqrt{s}=\sqrt{s_{NN}}$.  Some models of HI collisions predict
\cite{WG91, Li05, Ar961, Ar962} that particles produced in the forward
direction can have $p_L>\sqrt{s_{NN}}/2$ thereby violating Feynman
scaling assuming nucleon-nucleon $\sqrt{s}=\sqrt{s_{NN}}$
(Fig.~\ref{models}a).  Figure~\ref{models}b shows that HIJING
\cite{WG91} predicts progressively larger $p_L$ for produced particles
as the impact parameter ($b$), the distance between the colliding
nuclei, decreases.  It is noted in Ref.~\cite{Ar961, Ar962} that such
high-energy hadrons arise from interacting QCD strings.  They call
this a hadron accelerator, perhaps better as a QCD accelerator, since
HIJING predicts that mesons, meson pairs, and photons can have
$p_L>\sqrt{s_{NN}}/2$.  QCD string interactions are an example of the
collective behavior of partons.  However, such predictions are not well tested
by experiment.  Large-$x_F$ Drell-Yan production \cite{Co89} and
$J/\psi$ production \cite{Le00, An12} have been reported.  As well,
measurements of jet production in proton+lead collisions at $\sqrt{s_{NN}}$=5.02 TeV
were completed \cite{ATLAS14,CMS19}.  In all cases produced particles are observed with $x_F<1$.

As shown in Fig.~\ref{models}, HIJING predicts that the $p_L$ distribution
for particles produced in Cu+Au collisions at $\sqrt{s_{NN}}=200$ GeV
has similar high-energy behavior as production of these same particles
in p+p collisions at $\sqrt{s}=500$ GeV, or more.  These high-energy
particles are very relevant for calibrating high-energy cosmic-ray
detectors.  Models of HI collisions are used to determine the
composition of ultra-high energy cosmic rays \cite{Au14}, so checks of
models against collider data are important, as has been done by
comparing model predictions to LHC data \cite{En19}.  The EPOS model
\cite{Pi15}, which is widely used in measurements of high-energy cosmic
rays, respects Feynman scaling.  The Sibyll model \cite{Ri19} can not
be applied here since the target mass is limited.  The QGSJET model
\cite{Os11} respects Feynman scaling, and is closer to p+p at
$\sqrt{s}=200$ GeV since it does not include final-state interactions.
The DPMJET model \cite{En95} does violate Feynman scaling, although
the $p_L$ distribution does not extend as far as predicted by HIJING.
Due to the paucity of forward instrumentation at colliders, these
theoretical predictions of QCD string interactions are not tested by
measurement.  Generally, most theory work related to Feynman scaling
\cite{AMPS13, CGNS21} involves forward single-particle production.
However, jets produced in the forward direction are expected to
respect Feynman scaling as well \cite{KHM17}.  This article reports, for the
first time, relevant measurements to test theory predictions regarding
QCD string interactions.

We present measurements of yields of jets produced in the forward
direction from Cu+Au collisions at $\sqrt{s_{NN}}=200$ GeV.  The
measurements were made at interaction point (IP) 2 at the Relativistic
Heavy Ion Collider (RHIC) at Brookhaven National Laboratory in 2012.
The apparatus used for the measurements was previously discussed in
our report of forward-jet production in p+p collisions at
$\sqrt{s}=510$ GeV \cite{Bl15}.  The apparatus makes use of the
uniquely large ratio of insertion length (straight sections for
experimental apparatus) to $\sqrt{s}$ of RHIC.  The
Cu+Au collision data was obtained during a test of possible
pulse-shape discrimination in the hadron calorimeter, the
primary device used for jet finding.  Consequently,
there was no vernier scan \cite{Dr00} of the two beams meaning that cross sections
can not be measured.  We instead report fraction of minimum bias (MB)
as a yield measure.  The basic measurements are to find jets in the
hadron calorimeter that faces the Cu beam.  Jets are considered as a
function of $\Sigma Q_Y$, which is the charge sum in the beam-beam
counter (BBC) annulus that faces the Au beam.  Generally, $\Sigma
Q_Y<\Sigma Q_{Y,{\rm max}}$ is related to the impact parameter ($b$) of
the colliding ions through HI models, with smaller $b$ for larger
$\Sigma Q_Y$.  The apparatus, other details of the experiment,
the basic calibration of measuring devices, and aspects
of jet finding and corrections are discussed in the following section.

\section{Measurements and Analysis Methods}

The apparatus staged at IP2 was previously discussed \cite{Bl15}.  
A 200 cm $\times$ 120 cm forward calorimeter wall with a
central $(20~ {\rm cm})^2$ hole for the beams was constructed.  The
calorimeter wall was made from 236 cells.  Each cell was $117~ {\rm cm}
\times (10~ {\rm cm})^2$ of lead, with an embedded matrix of
$47\times47$ scintillating fibers that ran along the cell length in a
spaghetti calorimeter configuration \cite{Ar98}.  The calorimeter wall
faced the Cu beam and was positioned 530 cm from the IP.  The
calorimeter had $\approx$5.9 hadronic interaction lengths and $\approx$150
radiation lengths of material, so was ideal for finding jets.  The
calorimeter spanned the pseudorapidity range of $2.4<\eta<4.5$
for particles produced at the center of the vertex-$z$ ($z_v$) distribution.
Both charged and neutral particles follow straight-line trajectories
until they interact with matter since there was no analysis magnet.
The other primary components of the apparatus were two annular arrays
of 16 scintillator tiles, which served as beam-beam counters (BBC) \cite{Bi01}.
Each BBC array was positioned at $\pm150$ cm from the IP.  The
apparatus is modeled in GEANT \cite{GEANT}.

Triggering of event readout for this data was previously discussed
\cite{Bl15}.  Approximately half of the event sample is from a MB
trigger that required hits in both BBC annuli.  The other half of the
data sample is from a jet trigger that sums ADC values from symmetric
patches of the calorimeter to the left and to the right of the Cu
beam.  The jet trigger results can be emulated by applying the trigger
algorithm to MB data, thereby determining the equivalent number of MB
events for the jet trigger.  ADC data were pedestal-corrected and zero
suppressed.  Such data were acquired for the triggered bunch crossing,
and for the crossing before the trigger, and for the crossing after
the trigger.  The pre/post data were summed for towers associated with
jets and set a limit of $\approx0.001$ of the high-energy jets have
contributions from pileup, as expected because all components of the
detector apparatus were fast relative to the bunch-crossing frequency
and the average interaction rate was only $\approx10$ kHz.

The BBC reconstructed the $z$ component of the collision vertex ($z_v$) from
time-difference measurements, calibrated by the measured distance
between the two annuli.  Calibration to match arrival times of fast
particles produced in the collision was done online, with some final
adjustments done in offline analysis.  In the analysis, $|z_v|<75$ cm is
imposed.  The BBC is also used to
measure total charge from scintillation light.  The photomultiplier
tube gains were adjusted online to provide an average charge of 100
counts for minimum ionizing particles (MIPs) through one of the
detectors.  Final adjustments of the BBC charge calibration were done
in offline analysis.  The total charge measured in the Au-beam
facing BBC annulus ($\Sigma Q_Y$) is related by simulation to 
$b$ of the colliding ions.  HIJING 
simulations of minimum-bias Cu+Au collisions at $\sqrt{s_{NN}}=200$
GeV followed by GEANT simulations of the IP2 apparatus give a linear
dependence of $b$ with $\Sigma Q_Y$, with the
smallest $b\approx 9.6$ fm for a $\Sigma Q_Y<4000$ requirement.  The most
central ($b=0$) Cu+Au collisions are predicted to have $\Sigma Q_Y
\approx 30000$ counts.  For such collisions, the calorimeter is highly
occupied.  This report is restricted to peripheral Cu+Au collisions, to
ensure that jet clusters can be robustly found.

Most of the relevant calibration of the calorimeter was previously
described \cite{Bl15}, although corrections to the calibration are
required for Cu+Au collisions.  Peaks from MIPs from cosmic-ray muons
were matched to set the hardware gain of each cell prior to
collisions.  Software relative-gain corrections were made to match the
slopes of the steeply falling charge distributions for each cell from
p+p collision data to PYTHIA/GEANT simulations, which accurately
describe data from the individual calorimeter cells, and most other
aspects of the data.  These same gain
corrections work for the Cu+Au data, acquired two months later than
the p+p data, except for the 16 cells that had their anode signals
split by 50$\Omega$ splitters for the pulse-shape discrimination tests.  The signal split
meant twice higher software corrections (and energy range) for these
cells.  HIJING/GEANT simulations accurately describe the
individual-cell response to Cu+Au collisions, and many other aspects
of the data.  Following relative gain
calibration of all the cells, the absolute energy scale was determined
from reconstruction of neutral pions from pairs of photons detected in
the calorimeter.  The difference between the average hadronic and
electromagnetic energy scales was initially determined from
PYTHIA/GEANT simulations, and later confirmed by test-beam
measurements at FermiLab (T1064 \cite{T1064}).  T1064 also confirmed
good electromagnetic response of the calorimeter, as needed for
$\pi^0$ reconstruction.

After completing basic calibration of the calorimeter, jet finding
proceeds.  A jet finder is a pattern recognition algorithm \cite{Ca08} that
returns a list of jets having pseudorapidity ($\eta_{jet}^j$),
azimuthal angle ($\phi_{jet}^j$) and energy ($E_{jet}^j$) for
$j=1,...,M_{jet}$ from the list of $N$ hit elements of a calorimeter in
an event. The hit elements associated with one jet depend on a cone
size $R_{jet}=\sqrt{\Delta\eta^2+\Delta\phi^2}$ relative to the jet
direction.  The objects determined here are called {\it tower jets},
which are found either in data or in simulations of data using a
realistic event generator ({\it e.g.} PYTHIA \cite{PYTHIA}) and a
simulation of the IP2 apparatus using GEANT \cite{GEANT}.  Frequently, jets are
also found from the list of particles and their momenta returned by an event generator.
These are so-called {\it particle jets}, that do not have detector
effects and are built from particles of known energies.  The
jet-energy scale is set from comparison of tower jets and particle
jets, and checked in the data from 2- and 3-jet mass peaks.  The
resolution of the found jets has components from the detector
performance and from fluctuations of underlying event (UE) in Cu+Au
collisions, determined from embedding p+p jets from PYTHIA/GEANT into Cu+Au minimum-bias
data.  This section describes details of the
determination of the jet-energy scale and the resolution of the found
jets.

Jets are reconstructed using an implementation \cite{Bl15}
of the anti-$k_t$ algorithm \cite{Ca08} with a variety of cone radii
($R_{jet}$).  Checks of the jet finder were independently made by use
of the anti-$k_t$ option in the FastJet 3.3.2 package \cite{FJ11}. The
jet-energy scale was initially determined for $R_{jet}=0.7$ by
matching particle jets from PYTHIA \cite{PYTHIA} to tower jets from
PYTHIA/GEANT \cite{Bl15}, and then adjusted for other $R_{jet}$ values
by a linear correction made to the four momentum of each jet returned
by the jet finder.  This correction used the parton from PYTHIA/GEANT
simulations (where the energy of the parton is known) that was
associated with the reconstructed jet as a means of correlating jets
found with varying $R_{jet}$, and ensuring they have the same energy
for different $R_{jet}$.  We call this correction jet compensation.  A
requirement that a good jet has $|\eta_{jet}-3.25|<0.20$ is imposed to
ensure minimal effects from the limited acceptance of the calorimeter.
We establish that, on average, less than $\approx$3\% of the jets found
in Cu+Au events are fake, by randomly choosing towers from an ensemble
of events that had similar $z_v$ to within $\pm
5$cm and $\Sigma Q_Y$ to within $\pm 200$ counts, and then applying
the jet finder to events constructed from random towers.  The fake
jets fall more rapidly with increasing energy than real jets and also
have smaller tower multiplicity.

\begin{figure}
  \centering \includegraphics[height=2.0in]{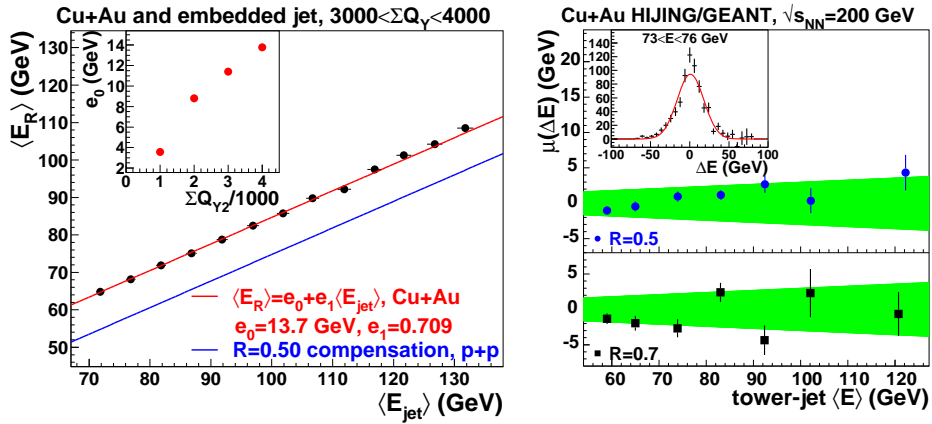} \caption{(left)
  Average reconstructed energy versus energy of the p+p jet embedded
  into minimum-bias Cu+Au data for events having $\Sigma Q_{Y1}<\Sigma
  Q_Y<\Sigma Q_{Y2}$.  The black points show the linear relationship
  between input and reconstructed jets.  The slope of the line fitting
  those points is the same as found for jet production in p+p.  The
  offset ($e_0$) represents the average effect of UE, which grows with
  $\Sigma Q_{Y2}$ as shown in the inset. (right) The jet-energy scale
  is determined from the comparison of particle jets ($E_P$) to tower
  jets ($E_R$) from HIJING/GEANT simulations of Cu+Au collisions.
  Results for $\Sigma Q_Y < 1000$ are shown.  The green band
  represents $\pm3$\% variation of the jet-energy scale.  The inset
  shows a distribution of energy difference $\Delta E = E_P-E_R$ from
  jet finding with $R_{jet}=0.5$ in one bin of particle-jet energy
  with a fit by a Gaussian distribution.  The blue points show the
  variation of Gaussian centroid ($\mu$) with the average tower-jet
  energy.  Results for jets found with $R_{jet}=0.5$ and $R_{jet}=0.7$
  are shown.}  \label{jes}
\end{figure}  

Underlying event (UE) is significantly larger in Cu+Au collisions than
in p+p collisions, and arises from particles that are not associated
with jets.  For Cu+Au data, a further correction to the jet-energy
scale is made from analysis of p+p jets from PYTHIA/GEANT embedded
into minimum-bias Cu+Au data.  Given the good agreement between p+p data
and PYTHIA/GEANT \cite{Bl15}, either data or simulation suffices for
embedding.  A distribution of the reconstructed jet energy results
when the reconstructed jet is directionally matched to the p+p jet
embedded into the event.  The mean value of the energy-difference
distribution from embedding at a given $\Sigma Q_Y$ is subtracted from
the jet energy.  Figure~\ref{jes} (left) shows that the embedding
results have the same slope with jet energy as found in p+p, with an
offset coming from the average effect of UE.  The inset to the figure
shows how the average additive correction to UE varies with $\Sigma
Q_Y$.  Hence, the jet-energy calibration is identical between p+p and
Cu+Au, except for a $\Sigma Q_Y$ additive correction needed for the
latter due to UE.  The impact of fluctuations of UE is discussed
below.

Multiple checks were made of the jet-energy scale for the Cu+Au data.
One check compares directionally matched particle jets of energy $E_P$
to tower jets of energy $E_R$ (Fig.~\ref{jes}) from HIJING/GEANT
simulations.  After correction for accidental matches caused by the
jet multiplicity, the distribution of the energy difference $\Delta E
= E_R-E_P$ is found to be Gaussian.  The Gaussian centroid is
consistent with being constant to within $\pm3$\% for the jet-energy
range $50 {\rm ~GeV}<E<130 {\rm ~GeV}$.  Figure~\ref{jes} shows results for events with
$\Sigma Q_Y<1000$.  Similar results are found for upper limits on
summed charge up to 4000 counts.  Other checks made of the jet-energy
scale are from mass peaks observed in multi-jet events, which will be
discussed in subsequent reports.

\begin{figure}
  \centering \includegraphics[height=3.0in]{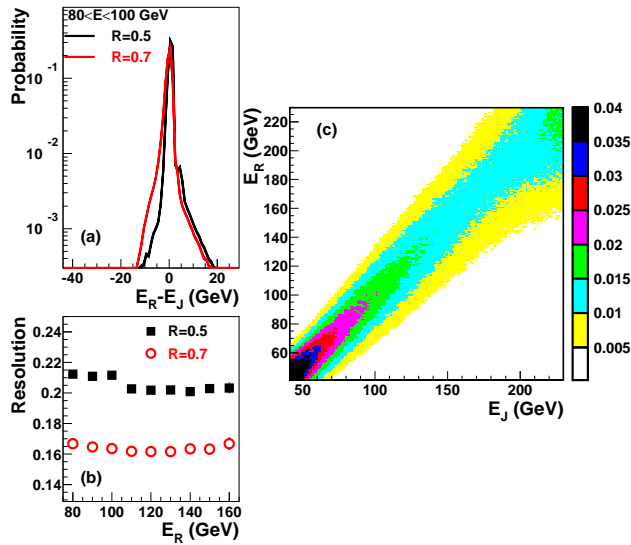} \caption{Jet
  resolution is determined from (a) embedding jets into Cu+Au MB
  events selected to have $\Sigma Q_Y<1000$.  The energy difference,
  $\Delta E$, between reconstructed and embedded jet is shown for one
  $E_J$ bin; (b) detector resolution is determined from comparing
  particle jets to tower jets.  The ratio $\sigma/\mu$ from Gaussian
  fits to energy difference distributions is shown versus tower-jet
  energy; (c) shows the product of the response matrix from embedding
  with the response matrix from detector resolution, normalized to
  give the probability a jet is reconstructed with $E_R$ when produced
  with energy $E_J$.}  \label{res}
\end{figure}

Jet-energy resolution must be quantified to unfold measured
spectra to recover the true spectrum generated in the collision.
There are two sources that affect jet resolution.  The
energy resolution of the calorimeter impacts the jet resolution.
Jet resolution can be worsened from particles that lie beyond the detector
acceptance ({\it e.g.}  soft photons from the decay of neutral pions
that fall beyond the jet cone, $R_{jet}$).  Jet resolution from
detector effects is determined from reconstructions of particle and
tower jets in PYTHIA/GEANT simulations of p+p collisions.  The GEANT
simulation is highly constrained since it accurately describes
test-beam data (T1064) \cite{T1064} for negative pions acquired at 6,
12, 18, and 24 GeV.  Figure~\ref{res}b shows that jet resolution from
detector effects changes little with jet energy, but does depend on
the cone radius used.  The resolution difference between $R_{jet}=0.5$
and 0.7 is due to differences in the ratio of electromagnetic particles
to hadronic particles within the jets.  This ratio is important
because T1064 data shows electron resolution is $\approx2.5$ times better than
negative pion resolution.  This ratio can be measured in principle,
but is unmeasured at present.  Consequently we attribute a systematic
uncertainty of half the difference between the $R_{jet}=0.5$ and 0.7
results for detector resolution, as discussed further below.

Another source of finite jet resolution for Cu+Au collisions is from
fluctuations of UE.  This resolution contribution is determined
from embedding p+p jets into minimum-bias Cu+Au, and then comparing the
reconstructed jet to the embedded jet.  Figure~\ref{res}a shows an
energy-difference distribution for events with $\Sigma Q_Y<1000$
counts.  There are exponential tails to the energy difference
distribution that fall from 1\% to 0.05\%.  These tails become more
prominent as $\Sigma Q_{Y,{\rm max}}$ is increased.  Figure~\ref{res}c shows
the response matrix obtained from embedding giving the probability to
observe a jet of energy $E_R$ from a jet produced with energy $E_J$.

\begin{figure}
  \centering
  \includegraphics[height=3.0in]{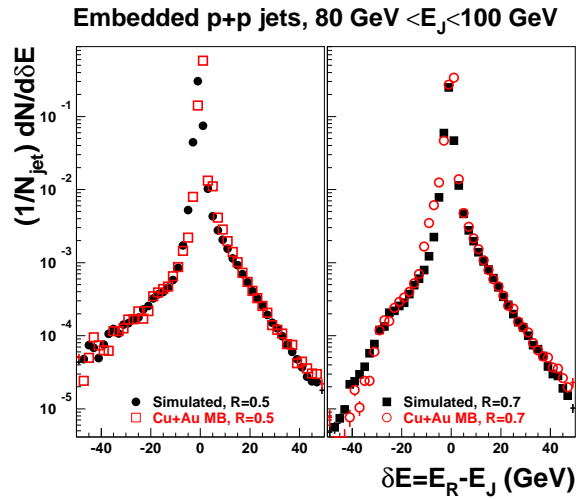}
  \caption{Comparison of results from embedding p+p jets into
  minimum-bias Cu+Au events (Fig.~\ref{res}a) to a 4-parameter
  simulation model described in the text.  This simulation model then
  has parameters adjusted to estimate the impact of UE fluctuations on
  the results.}
  \label{sim}
\end{figure}  

A simulation model was developed to explain embedding results
(Fig.~\ref{res}a).  The objective is to use the model to explore
systematic uncertainties associated with UE.  Fig.~\ref{sim} shows
that the embedding results are well explained by a simple model.  In
the model a p+p jet of energy $E_J$ is selected.  Additional towers
are randomly produced.  Four parameters are introduced to describe the
two-component tower energy distributions from minimum-bias Cu+Au
events, which the simulation model can reproduce as can HIJING/GEANT
simulations.  One component of the tower energy distribution is a
Gaussian distribution whose centroid is zero.  The probability ($P_G$)
and $\sigma$ are two parameters of the model, with $\sigma=35/r$ GeV,
where $r$ is the distance of the tower center from the
beam which varies from 5 cm to 116 cm.  The $P_G$ parameter is 0.02,
and represents the probability a cell becomes a tower with energy
drawn at random from the Gaussian distribution.  The second component
of the tower-energy distribution is exponential.  This component is
modeled by two parameters corresponding to the probability,
$P=1.2\times10^{-3}$, that a cell is a tower whose energy is drawn
at random from an exponential distribution specified by slope
$E_T=180/r$ GeV.  The $P$ parameter increases quadratically with
$\Sigma Q_{Y,{\rm max}}$.  For $\Sigma Q_{Y,{\rm max}}=1000$, $P=4.2\times10^{-3}$
for $R=0.5$ jets and $P=2.4\times10^{-3}$ for $R=0.7$ jets. This
4-parameter simulation model accurately describes the embedding
results.  Systematic uncertainties from UE are estimated by scaling
the probabilities by a factor of two, discussed further below.

\begin{figure}
  \centering
  \includegraphics[height=2.9in]{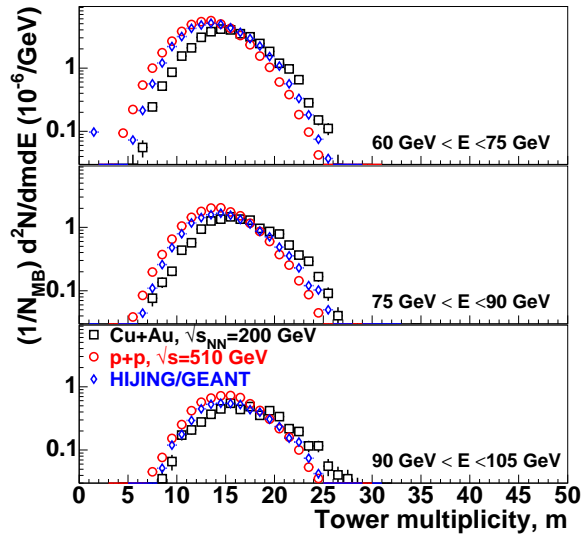}
  \caption{Multiplicity of towers within jets found in Cu+Au data, in
    HIJING/GEANT simulations, and in p+p collisions at $\sqrt{s}=510$ GeV.}
  \label{mult}
\end{figure}  

\section{Results}

The multiplicity of towers found in jets from Cu+Au collisions is
shown in Fig.~\ref{mult}.  Comparison is made to HIJING/GEANT
simulations.  Although the extent of the data and simulation
multiplicity distributions match for all energy bins, the average
multiplicity from simulation is one tower less than found in the
data.  Also, the Cu+Au data are compared to jets from
p+p at $\sqrt{s}=510$ GeV, as motivated by Fig.~\ref{models}a.  The mean tower multiplicity is two towers
smaller in p+p compared to Cu+Au, most likely due to UE
contributions in HI data.  The jet energies are corrected as per
Fig.~\ref{jes} (left); the tower multiplicities are not corrected.  Trends in the
comparisons persist from low to high energy, with energy independent
scale factors used for HIJING/GEANT and p+p.  From this we conclude
jets with $E>90$ GeV are not from the merging of two or more lower energy jets.
This conclusion is supported by embedding studies, as well.

\href{https://www.andy.bnl.gov/cuau/files/210514}{Single event displays}
show that most events with high-energy jets from peripheral Cu+Au
collisions are relatively simple, similar to what is observed for p+p collisions,
consistent with embedding results from Fig.~\ref{sim} that show that
most p+p jets are unaffected by UE.
There are some events with a high energy jet that also have a large value
for the energy sum of cells closest to the beam (the perimeter-1 sum,
$\Sigma E_{P1}$), that make the jet cluster less distinct from other
activity in the calorimeter.  We limit any effect of
such events by imposing $\Sigma E_{P1}<350$ GeV as an event requirement.
Absent this restriction, jets are found to even higher energy than
what we report.  The
restriction is imposed to ensure minimal confusion for the jet
finder.  For MB events, the distribution of $\Sigma E_{P1}$ is similar
for different $\Sigma Q_Y$ bins considered.  
HIJING/GEANT predicts a similar $\Sigma E_{P1}$ distribution to that
seen in the data, precluding single-beam backgrounds as the source of
these events.

Jet-energy resolution impacts the steeply falling energy
distributions.  The small yield of high-energy jets may be caused by
migration of some of the prolific lower-energy jets to high energy,
due to finite resolution.  

Both detector resolution and UE fluctuations are corrected for in
the data by the singular value decomposition (SVD) unfolding
algorithm \cite{HK95}.  One input to the SVD unfolding algorithm is a normalized
response matrix, shown in Fig.~\ref{res}c for events with $\Sigma
Q_Y<1000$ counts.  The color scale represents the probability to
reconstruct a jet with energy $E_R$ from jets produced with energy
$E_J$.  The response matrix is a product of a matrix obtained from
embedding studies with a matrix representing detector resolution.
The observed jet-energy distribution equals the full normalized
response matrix times the true $dN/dE$, represented as a vector.

\begin{figure}
  \centering
  \includegraphics[height=3.0in]{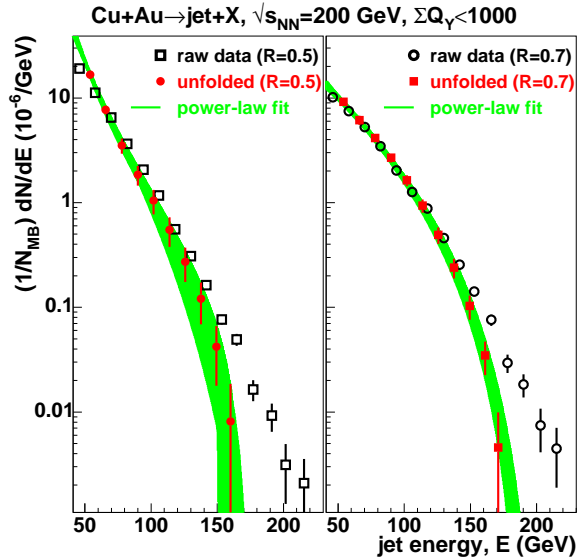}
  \caption{Jet-energy distribution from Cu+Au collisions with $\Sigma
    Q_Y<1000$.  Open points are raw data for jets found with $R_{jet}=0.5$
    (left) and $R_{jet}=0.7$ (right).  Solid points are from unfolding
    \cite{HK95} the response matrix (Fig.~\ref{res}c) from the
    data.  The unfolded points are fit by power-law functions whose
    endpoints ($E_0$) are $>100$ GeV.  At fixed $E$, the width of the
    green band is predominantly from the uncertainty in the jet-energy
    scale.  The power-law parameters are listed in Table~\ref{fitpar}}
  \label{dnde-sq1}
\end{figure}

The SVD unfolding algorithm \cite{HK95} was extensively studied using
power-law parameterizations of jet-energy spectra: $dN/dE = N
(1-x)^p/x^q$, with $x=E/E_0$ identified as the Feynman scaling
variable.  A description of the unfolding tests is available in an
analysis note \cite{AnaNote}.  As the impact parameter of the colliding ions decreases, the
jet-energy spectra are best described by an exponential function,
which from tests may not be uniquely unfolded to eliminate resolution
effects and return the true distribution of jets from the collision.
For more peripheral HI collisions, negative logarithmic curvature is
observed in raw jet-energy distributions (Fig.~\ref{dnde-sq1}).  The
raw data extends to energies that are more than two times larger than
given by Feynman scaling assuming $\sqrt{s}=\sqrt{s_{NN}}$.  The
unfolding proceeds via a singular value decomposition \cite{HK95} of
the response matrix (Fig.~\ref{res}c).  Instabilities inherent to SVD
unfolding are cured by regularizing the singular values.  The
statistics per bin of the response matrix are sufficient to ensure
stable unfolding results.  The regularization parameter, $\tau$ is
determined by computing a $\chi^2$ between the unfolded distribution
and an input guide distribution that is initially determined from a
power-law fit to the raw data, with the power-law parameters $N, p, q,
E_0$ allowed to freely vary.  Other choices for the regularization
parameter were also made with similar results, albeit with some change
in $E_0$ accounted for by including a systematic uncertainty from the
regularization parameter.  Figure~\ref{dnde-sq1} shows that finite
resolution does not account for all of the highest energy jets, in
that unfolded data can be fit by power-law functions with
$E_0=167.0\pm10.8$ GeV when $R_{jet}=0.5$ is used and
$E_0=197.5\pm7.6$ GeV when $R_{jet}=0.7$ is used.  The unfolded
results are similar to an independent Monte Carlo method that adjusts
the true jet $dN/dE$ until the response matrix acting on it produces
the observed jet-energy distribution.  The fitted $E_0$ values
correspond to effective NN $\sqrt{s}\approx$ 334 and 395 GeV, which is
significantly larger than the assumption that nucleon-nucleon
$\sqrt{s}=\sqrt{s_{NN}}$, similar to model expectations
(Fig.~\ref{models}a).  The $E_0$ values are determined from the
average of six power-law fits obtained as the jet-energy scale, the
jet-energy resolution, the SVD regularization parameter, and UE
fluctuations are varied.  The uncertainties on $E_0$ are the root-mean
squared values from the distributions.  The yields of jets are similar
when $R_{jet}=0.5$ and $R_{jet}=0.7$ are used to find jets.  The
smaller cone size results in a steeper rise of the jet-energy spectrum
with decreasing energy at low energy, most likely due to splitting of
higher energy jets.

The unfolded data in Fig.~\ref{dnde-sq1} includes multiple sources of
systematic uncertainty added in quadrature with the statistical
uncertainty.  The $\pm$3\% jet-energy scale uncertainty is included in
both the raw-data and unfolded points.  The width of the power-law fit
parameter band is predominantly from jet-energy scale variations.
Jet-energy resolution is also included in the uncertainties for the
unfolded data points.  Half the difference between the R=0.5 and R=0.7
jet-energy resolutions (Fig.~\ref{res}b) is considered the uncertainty
and separate SVD unfoldings are performed for the nominal and modified
jet-energy resolution.  Variation of the SVD unfolding parameter
$\tau$ also contributes to the systematic uncertainty.  Finally, the
simulation that describes embedding results (Fig.~\ref{sim}) has the
tower probability from both Gaussian and exponential distributions
artificially increased by a factor of two.  Although this may appear
a large variation, it illustrates that the dominant effects from
unfolding are from detector effects, both jet-energy scale and
jet-energy resolution.  The impact of UE fluctuations is small, as can
be seen from the small probability of exponential tails in
Fig.~\ref{sim}.  Systematic effects from jet-energy resolution can be
reduced in an experiment that measures the electromagnetic fraction of
the jet.

\begin{table}[h]
 \begin{center}
  \begin{tabular}{ | l | l | l |}
   \hline
   ~~~~Parameter~~~~&~~~~$R=0.5$~~~~&~~~~$R=0.7$~~~    \\ \hline
   ~~~$E_0$ (GeV)~~~&~$167.0\pm10.8$~~&~$197.5\pm7.6$~ \\ \hline
   ~~~~~$N$~~~~~    &~$293\pm191$~&~$2240\pm411$~  \\ \hline
   ~~~~~$p$~~~~~    &~$1.9\pm0.6$~&~$3.8\pm0.3$~   \\ \hline
   ~~~~~$q$~~~~~    &~$3.0\pm0.3$~&~$0.4\pm0.1$~   \\ \hline
  \end{tabular}
  \caption{Average values of power-law fit parameters in
  Fig.~\ref{dnde-sq1}.  Individual power-law fits to unfolded data are
  performed during systematic variations of jet-energy scale,
  jet-energy resolution, the SVD reguarlization parameter, and UE, as
  described in the text. \label{fitpar}}
 \end{center}
 
\end{table}

Table~\ref{fitpar} lists the average value of the power-law fit
parameters from the variations used to deduce systematic uncertainty
in Fig.~\ref{dnde-sq1}.  The normalization parameter $N$ is strongly
correlated with the fitted powers $p,q$.  We attribute physical
significance to only the $E_0$ parameter, although future
extensions of HI models to accommodate a QCD accelerator may find such
significance in the other parameters.

Run-by-run analysis shows that no special beam
conditions give rise to these high-energy jets.  The $p_T$
distribution of the highest-energy tower within the jet qualitatively
matches predictions from HIJING/GEANT, again precluding single-beam
backgrounds as the origin of the high-energy jets.  Similar yields of
high-energy jets are found from different parts of the $z_v$
distribution.  Contributions from
pileup are found to be small.  We rule out instrumental or environmental sources of
these high-energy jets.

After all corrections and systematic checks, we find that forward-jet
production in peripheral Cu+Au collisions exceeds the Feynman-scaling endpoint energy
by a factor between 1.7 and 2.0, assuming conventional parton scattering from
nucleon-nucleon collisions with $\sqrt{s}=\sqrt{s_{NN}}$.

\begin{figure}
  \centering
  \includegraphics[height=3.0in]{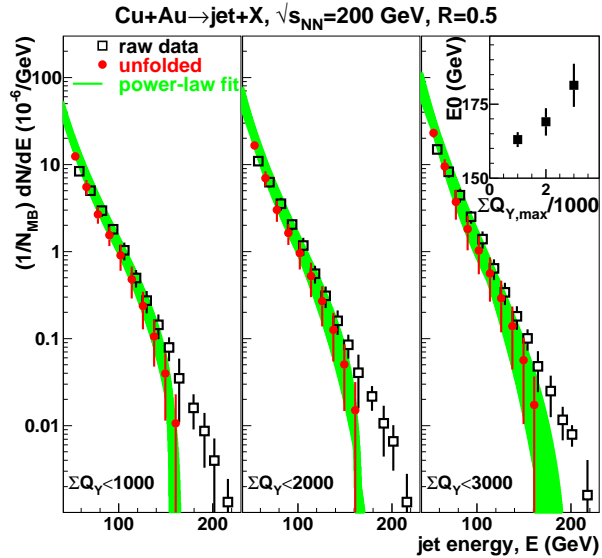}
  \caption{Jet-energy distribution from Cu+Au collisions as a function
    of the upper limit on $\Sigma Q_Y$.  Open points are raw data
    for jets found with $R_{jet}=0.5$ .  Solid points are from unfolding
    \cite{HK95} the response matrix (Fig.~\ref{res}) from the
    data.  The inset shows the fitted endpoint ($E_0$) as a function
    of the upper limit on $\Sigma Q_Y$.  Uncertainties on $E_0$ are
    the quadrature sum of uncertainties from power-law fits and
    systematic effects from detector-based jet resolution.}
  \label{unfold-sumqy}
\end{figure}  

\begin{table}[h]
 \begin{center}
  \begin{tabular}{ | l | l | l | l |}
   \hline
   ~~~~Parameter~~~~&~$\Sigma Q_{Y,{\max}}$=1000~&~$\Sigma Q_{Y,{\max}}$=2000~&~$\Sigma Q_{Y,{\max}}$=3000~\\ \hline
   ~~~$E_0$ (GeV)~~~&~$163.0\pm1.9$~&~$169.0\pm4.4$~&~$181.3\pm7.2$~\\ \hline
   ~~~~~$N$~~~~~    &~$87\pm17$~&~$44\pm9$~&~$39\pm9$~\\ \hline
   ~~~~~$p$~~~~~    &~$1.3\pm0.2$~&~$1.5\pm0.5$~&~$1.2\pm0.3$~\\ \hline
   ~~~~~$q$~~~~~    &~$3.1\pm0.1$~&~$3.5\pm0.3$~&~$3.7\pm0.5$~\\ \hline
  \end{tabular}
  \caption{Average values of power-law fit parameters in
  Fig.~\ref{unfold-sumqy}.  Individual power-law fits to unfolded data are
  performed during systematic variations of jet-energy scale,
  jet-energy resolution, the SVD reguarlization parameter, and UE, as
  described in the text. \label{fit-sumqy}} 
 \end{center}
\end{table}

The HIJING model of heavy-ion collisions \cite{WG91} predicts that
single-particle production also violates Feynman scaling, with the
magnitude of the effect increasing as the collisions become more
central ($b\rightarrow0$).  To examine this, we systematically
increase the upper limit on the Au-beam facing charge sum, as shown in
Fig.~\ref{unfold-sumqy}.  Checks of the data show that multi-jet
mass-peak widths are proportional to the root-mean square of the
embedding $\Delta E$ distribution (Fig.~\ref{res}a) as $\Sigma Q_{Y,{\rm
max}}$ is increased.  This supports that embedding properly measures
resolution from UE fluctuations. To ensure adequate statistics, the
acceptance is increased to $|d\eta|<0.25$.  Systematic uncertainties are
estimated by varying the jet-energy scale by $\pm3$\%, the jet-energy
resolution by $\pm3$\%, the SVD regularization parameter, and UE
fluctuations by a factor of 1.5.  The uncertainties shown are the
quadrature sum of statistical uncertainty and systematic uncertainties
from the four variations considered.  The jet-energy scale variations
makes the largest contribution to the systematic uncertainty, although
the SVD regularization variation is dominant for high-energy jets when
$\Sigma Q_{Y,{\rm max}}$=3000.  As $\Sigma Q_{Y,{\rm max}}$ is increased
the fitted $E_0$ value increases, as shown in the inset to
Fig.~\ref{unfold-sumqy}.  This is consistent with expectations
regarding string interactions becoming a more effective QCD
accelerator at smaller impact parameter.  At the very least,
Fig.~\ref{unfold-sumqy} demonstrates that $E_0>\sqrt{s_{NN}}/2$ is not
resulting from some special centrality selection.  Smooth dependence
on $\Sigma Q_{Y{\rm max}}$ is observed in all cases.  Unlike for HIJING, even the
most peripheral Cu+Au collisions still have $E_0>100$ GeV in unfolded
data.  The average values of the other fit parameters are shown in
Table~\ref{fit-sumqy}.  QCD string interactions have been in some HI
collision models for over thirty years.  These data provide clear
experimental evidence of QCD string interactions.  Models will have to
be adjusted to better describe these findings.  Given the complexity
of Cu+Au collisions, it could be possible that some combination of
initial-state effects ({\it i.e.} modification of parton distributions
in nuclei) and final-state effects could explain our measurements.

\section{Conclusions}

Jets produced by Cu+Au collisions at $\sqrt{s_{NN}}=200$ GeV in the
forward direction significantly violate Feynman scaling, assuming
nucleon-nucleon $\sqrt{s}=\sqrt{s_{NN}}$.  The effective
nucleon-nucleon $\sqrt{s}$ varies linearly from 1.5 to $\approx2$ times
larger than $\sqrt{s_{NN}}$ as the impact parameter decreases.  These
observations qualitatively match previously untested theoretical
expectations of QCD string interactions that are within multiple
models of HI collisions.  There are important implications of QCD
string interactions for the study of very high-energy cosmic rays,
particularly for observables that rely on models of the interaction of
cosmic-ray primaries in the atmosphere.  It is fully expected that QCD
string interactions may also provide a window to search for the
production of new particles.

\begin{acknowledgements}
We thank the RHIC Operations Group at BNL.  This work was supported in
part by the Office of NP within the U.S. DOE Office of Science, the Ministry
of Ed. and Sci. of the Russian Federation, and the Ministry of Sci., Ed.
and Sports of the Rep. of Croatia, and IKERBASQUE and the UPV/EHU.  We also
thank the operations group and the test beam staff at FNAL.
\end{acknowledgements}

\raggedright

\vspace{10mm}



\begin{thebibliography}{10}

\expandafter\ifx\csname url\endcsname\relax
  \def\url#1{\texttt{#1}}\fi
\expandafter\ifx\csname urlprefix\endcsname\relax\def\urlprefix{URL~~}\fi

\bibitem{Ar961}
  \bibinfo{author}{N.~Armesto,} \bibinfo{author}{M.~A.~Braun,}
  \bibinfo{author}{E.~G.~Ferreiro,} \bibinfo{author}{C.~Pajares,}
  \bibinfo{author}{and Yu.~M.~Shabelski,}
  \newblock \href{https://www.sciencedirect.com/science/article/abs/pii/S0370269396012488?via%3Dihub}
  {\bibinfo{journal}{Phys. Lett.}
  \textbf{\bibinfo{volume}{B389}}, \bibinfo{pages}{78-82} (\bibinfo{year}{1996})}.

\bibitem{Ar962}
  \bibinfo{author}{N.~Armesto,} \bibinfo{author}{M.~A.~Braun,}
  \bibinfo{author}{E.~G.~Ferreiro,} \bibinfo{author}{C.~Pajares,}
  \bibinfo{author}{and Yu.~M.~Shabelski,}
  \href{https://arxiv.org/abs/hep-ph/9606333}{https://arxiv.org/abs/hep-ph/9606333}.
    
\bibitem{STAR02}
  \bibinfo{author}{STAR collaboration, }\bibinfo{author}{C.~Adler} \emph{et al.,}
  \newblock
  \href{https://doi.org/10.1103/PhysRevLett.89.202301}
  {\bibinfo{Journal}{Phys. Rev. Lett.}
  \textbf{\bibinfo{volume}{89}}, \bibinfo{pages}{202301} (\bibinfo{year}{2002})}.

\bibitem{PHEN02}
  \bibinfo{author}{PHENIX
  collaboration, }\bibinfo{author}{K.~Adcox} \emph{et al.,}
  \newblock
  \href{https://doi.org/10.1103/PhysRevLett.88.022301}
  {\bibinfo{Journal}{Phys. Rev. Lett.}
  \textbf{\bibinfo{volume}{88}}, \bibinfo{pages}{022301}
  (\bibinfo{year}{2001})}.

\bibitem{CMS12}
  \bibinfo{author}{CMS collaboration, }
  \newblock
  \href{https://doi.org/10.1016/j.physletb.2012.04.058}
  {\bibinfo{Journal}{Phys. Lett. B}
  \textbf{\bibinfo{volume}{712}}, \bibinfo{pages}{176}
  (\bibinfo{year}{2012})}.

\bibitem{ATLAS13}
  \bibinfo{author}{ATLAS collaboration, }
  \newblock
  \href{https://doi.org/10.1016/j.physletb.2013.01.024}
  {\bibinfo{Journal}{Phys. Lett. B}
  \textbf{\bibinfo{volume}{719}}, \bibinfo{pages}{220}
  (\bibinfo{year}{2013})}.

\bibitem{ALICE15}
  \bibinfo{author}{ALICE collaboration, }
  \newblock
  \href{https://doi.org/10.1016/j.physletb.2015.04.039}
  {\bibinfo{Journal}{Phys. Lett. B}
  \textbf{\bibinfo{volume}{746}}, \bibinfo{pages}{1}
  (\bibinfo{year}{2015})}.

\bibitem{BRS18}
  \bibinfo{author}{W.~Busza,} \bibinfo{author}{K.~Rajagopal,} \bibinfo{author}{and
  W.~van der Schee}
  \newblock
  \href{https://doi.org/10.1016/j.physletb.2013.01.024}
  {\bibinfo{Journal}{Ann. Rev. Nucl. Part. Sci.}
  \textbf{\bibinfo{volume}{68}}, \bibinfo{pages}{339} (\bibinfo{year}{2018})}.

\bibitem{Fe69}
  \bibinfo{author}{R.~Feynman,}
  \href{https://doi.org/10.1103/PhysRevLett.23.1415}
  {\newblock \bibinfo{Journal}{Phys. Rev. Lett.}
  \textbf{\bibinfo{volume}{23}}, \bibinfo{pages}{1415} (\bibinfo{year}{1969})}.
  
\bibitem{WG91}
  \bibinfo{author}{X.~N.~Wang} \bibinfo{author}{and M.~Gyulassy,}
  \newblock
  \href{https://journals.aps.org/prd/abstract/10.1103/PhysRevD.44.3501}
  {\bibinfo{journal}{Phys. Rev.}
  \textbf{\bibinfo{volume}{D44}}, \bibinfo{pages}{3501}
  (\bibinfo{year}{1991})}.


\bibitem{Co89}
  \bibinfo{author}{J.~S.~Conway} \emph{et~al.,}
  \newblock
  \href{https://doi.org/10.1103/PhysRevD.39.92}
  {\bibinfo{journal}{Phys. Rev.}
  \textbf{\bibinfo{volume}{D 39}}, \bibinfo{pages}{92}
  (\bibinfo{year}{1989})}.


\bibitem{Le00}
  \bibinfo{author}{M.~J.~Leitch, }
  \bibinfo{author}{W.~M.~Lee, }
  \bibinfo{author}{M.~E.~Beddo, }
  \bibinfo{author}{C.~N.~Brown, }
  \bibinfo{author}{T.~A.~Carey, }
  \bibinfo{author}{T.~H.~Chang, } \emph{et~al.,}
  \newblock
  \href{https://doi.org/10.1103/PhysRevLett.84.3256}
  {\bibinfo{journal}{Phys. Rev. Lett.}
  \textbf{\bibinfo{volume}{84}}, \bibinfo{pages}{3256}
  (\bibinfo{year}{2000})}.

\bibitem{An12}
  \bibinfo{author}{R.~Analdi} \emph{et al.,}
  \newblock
  \href{https://www.sciencedirect.com/science/article/abs/pii/S0370269311014146?via%3Dihub}
  {\bibinfo{journal}{Phys. Lett. B}
  \textbf{\bibinfo{volume}{706}}, \bibinfo{pages}{263}
  (\bibinfo{year}{2012})}.


\bibitem{ATLAS14}
  \bibinfo{author}{ATLAS collaboration, }
  \newblock
  \href{https://doi.org/10.1016/j.physletb.2015.07.023}
  {\bibinfo{journal}{Phys. Lett. B}
  \textbf{\bibinfo{volume}{748}}, \bibinfo{pages}{392}
  (\bibinfo{year}{2015})}.

\bibitem{CMS19}
  \bibinfo{author}{CMS collaboration, }
  \newblock
  \href{https://doi.org/10.1007/JHEP05%282019%29043}
  {\bibinfo{journal}{JHEP} \textbf{\bibinfo{volume}{05}}, \bibinfo{pages}{043}
  (\bibinfo{year}{2019})}.

\bibitem{Li05}
  \bibinfo{author}{Zi-Wei~Lin,}
  \bibinfo{author}{Che Ming Ko,}
  \bibinfo{author}{Bao-An~Li,}
  \bibinfo{author}{Bin~Zhang,}
  \bibinfo{author}{and Subrata~Pal}
  \newblock
  \href{https://journals.aps.org/prc/abstract/10.1103/PhysRevC.72.064901}
  {\bibinfo{journal}{Phys. Rev.}
  \textbf{\bibinfo{volume}{C 72}}, \bibinfo{pages}{064901}
  (\bibinfo{year}{2005})}.

\bibitem{Au14} \bibinfo{author}{A.~Aab}  \emph{et~al.,}
  \newblock
  \href{https://doi.org/10.1103/PhysRevD.90.122006}
  {\bibinfo{journal}{Phys. Rev. D} \textbf{\bibinfo{volume}{90}},
    \bibinfo{pages}{122006} (\bibinfo{year}{2014})}.

\bibitem{En19}
  \bibinfo{author}{D.~d'Enterria}
  \bibinfo{author}{and T.~ Pierog,}
  \newblock
  \href{https://link.springer.com/article/10.1007/JHEP08%282016%29170}
  {\bibinfo{journal}{JHEP}
  \textbf{\bibinfo{volume}{1608}} \bibinfo{pages}{170}
  (\bibinfo{year}{2016})}.
    
\bibitem{Pi15}
  \bibinfo{author}{T.~Pierog,}
  \bibinfo{author}{Iu.~Karpenko,}
  \bibinfo{author}{J.M.~Katzy,}
  \bibinfo{author}{E.~Yatsenko,}
  \bibinfo{author}{and K.~Werner}
  \newblock
  \href{https://journals.aps.org/prc/abstract/10.1103/PhysRevC.92.034906}
  {\bibinfo{journal}{Phys. Rev.}
  \textbf{\bibinfo{volume}{C 92}}, \bibinfo{pages}{034906}
  (\bibinfo{year}{2015})}.

\bibitem{Ri19}
  \bibinfo{author}{F.~Riehn,}
  \bibinfo{author}{R.~Engel,}
  \bibinfo{author}{A.~Fedynitch,}
  \bibinfo{author}{T.~K.~Gaisser,}
  \bibinfo{author}{and T.~ Stanev,}
  \newblock
  \href{https://doi.org/10.1103/PhysRevD.102.063002}
  {\bibinfo{journal}{Phys. Rev. D} \textbf{\bibinfo{volume}{102}}
  \bibinfo{pages}{063002} (\bibinfo{year}{2020})}.
  
\bibitem{Os11}
  \bibinfo{author}{S.~Ostapchenko,}
  \newblock
  \href{https://doi.org/10.1103/PhysRevD.83.014018}
  {\bibinfo{journal}{Phys. Rev. D} \textbf{\bibinfo{volume}{83}},
  \bibinfo{pages}{014018} (\bibinfo{year}{2011})}.
  
\bibitem{En95}
  \bibinfo{author}{R.~Engel,}
  \newblock
  \href{http://inspirehep.net/record/373339}
  {\bibinfo{journal}{Z. Phys.} \textbf{\bibinfo{volume}{C66}},
  \bibinfo{pages}{203-214} (\bibinfo{year}{1995})}.
  
\bibitem{AMPS13}
  \bibinfo{author}{G.~H.~Arakelyan,} \bibinfo{author}{C.~Merino,}
  \bibinfo{author}{C.~Pajares,} \bibinfo{author} {and Yu.~M.~Shabelski,}
  \newblock
  \href{https://link.springer.com/article/10.1134%2FS1063778813020026}
  {\bibinfo{Journal}{Phys. Atom. Nucl.}
  \textbf{\bibinfo{volume}{76}}, \bibinfo{pages}{316-325}
  (\bibinfo{year}{2013})}.

\bibitem{CGNS21}
  \bibinfo{author}{F.~Carvalho,} \bibinfo{author}{V.~P.~Goncalves, }
  \bibinfo{author}{F.~S.~Navarra, } \bibinfo{author}{and D.~Spiering, }
  \newblock
  \href{https://journals.aps.org/prd/abstract/10.1103/PhysRevD.103.034021}
  {\bibinfo{Journal}{Phys. Rev. D}
  \textbf{\bibinfo{volume}{103}}, \bibinfo{pages}{034021}
  (\bibinfo{year}{2021})}.

\bibitem{KHM17}
  \bibinfo{author}{K.~Kutak, } \bibinfo{author}{H.~V.~Haevermaet, }
  \bibinfo{author}{and P.~V.~Mechelen, }
  \newblock
  \href{https://www.sciencedirect.com/science/article/pii/S0370269317303593?via%3Dihub}
  \bibinfo{Journal}{Phys. Lett. B}
  \textbf{\bibinfo{volume}{770}}, \bibinfo{pages}{412-417}
  (\bibinfo{year}{2017}).

\bibitem{Bl15}
  \bibinfo{author}{L.~C.~Bland} \emph{et~al.,}
  \newblock
  \href{https://www.sciencedirect.com/science/article/pii/S0370269315007522?via%3Dihub}
  {\bibinfo{Journal}{Phys. Lett.}
  \textbf{\bibinfo{volume}{B750}}, \bibinfo{pages}{660-665}
    (\bibinfo{year}{2015})}.

\bibitem{Dr00}
  \bibinfo{author}{A.~Drees,}
  \newblock
  \href{https://www.osti.gov/servlets/purl/784218}
  {\bibinfo{Journal}{BNL report} \bibinfo{volume}{67961} (\bibinfo{year}{2000})}.

\bibitem{Ar98}
  \bibinfo{author}{T.~A.~Armstrong} \emph{et~al.,}
  \newblock
  \href{https://doi.org/10.1016/S0168-9002(98)91984-2}
  {\bibinfo{Journal}{Nucl. Instrm. Meth.}
  \textbf{\bibinfo{volume}{A 406}}, \bibinfo{pages}{227-258}
  (\bibinfo{year}{1998})}.

\bibitem{Bi01}
  \bibinfo{author}{R.~Bindel} \emph{et~al.,}
  \newblock
  \href{https://doi.org/10.1016/S0168-9002(01)00866-X}
  {\bibinfo{Journal}{Nucl. Instrm. Meth.}
  \textbf{\bibinfo{volume}{A 474}}, \bibinfo{pages}{38-45}
  (\bibinfo{year}{2001})}.
  
\bibitem{Ca08}
  \bibinfo{author}{M.~Cacciari, } \bibinfo{author}{G.~P.~Salam, }
  \bibinfo{author}{G.~Soyez, }
  \newblock
  \href{https://iopscience.iop.org/article/10.1088/1126-6708/2008/04/063}
  {\bibinfo{journal}{JHEP}
  \textbf{\bibinfo{volume}{2008}}, \bibinfo{pages}{063} (\bibinfo{year}{2008})}.

\bibitem{FJ11}
  \bibinfo{author}{M.~Cacciari, } \bibinfo{author}{G.~P.~Salam, }
  \bibinfo{author}{and G.~Soyez, }
  \newblock
  \href{https://link.springer.com/article/10.1140/epjc/s10052-012-1896-2}
  {\bibinfo{journal}{Eur. Phys. Journ.}
  \textbf{\bibinfo{volume}{72}}, \bibinfo{pages}{1896}
  (\bibinfo{year}{2012})}.

\bibitem{PYTHIA}
  \bibinfo{author}{T.~Sj\"ostrand, } \bibinfo{author}{S.~Mrenna, } \bibinfo{author}{and P.~Skands, }
  \newblock
  \href{https://iopscience.iop.org/article/10.1088/1126-6708/2006/05/026}
  {\bibinfo{journal}{JHEP}
  \textbf{\bibinfo{volume}{2006}} (\bibinfo{year}{2006})}.

\bibitem{T1064}
  \bibinfo{author}{H.~J.~Crawford, }
  \href{https://www.andy.bnl.gov/cuau/files//T1064\_tsw.pdf}{
  \newblock \bibinfo{title}{\it T1064, STAR Forward Calorimeter System,
  Technical Scope of Work for the 2015 FermiLab Test Beam Facility Program}}.


\bibitem{HK95}
\bibinfo{author}{A.~H\"ocker } \bibinfo{author}{and V.~Kartvelishvili, }
\newblock
\href{https://www.sciencedirect.com/science/article/abs/pii/0168900295014780?via%3Dihub}
{\bibinfo{journal}{Nucl. Inst. Meth.} \textbf{\bibinfo{volume}{372}},
  \bibinfo{pages}{469-481} (\bibinfo{year}{1995})}.

\bibitem{AnaNote}
\bibinfo{author}{L.~C.~Bland, }
\bibinfo{author}{H.~J.~Crawford, }
\bibinfo{author}{A.~Quintero}
\href{https://www.andy.bnl.gov/cuau/files/analysis-note-final.pdf}
{\bibinfo{title}{{\it Forward Dijets at RHIC:  Relevant Details},}}
\bibinfo{year}{2022}
\bibinfo{Journal}{(unpublished)}.

\bibitem{GEANT}
  \bibinfo{author}{Ren\'e Brun} \emph{et~al.,}
  \newblock \bibinfo{Journal}{CERN-W-5013}
  (\bibinfo{year}{1994}).
  
\end{thebibliography}
\end{document}